E. Nikitina, L. Urazova, O. Churuksaeva
DINAMICS OF HPV INFECTION AMONG WOMEN WITH CERVICAL LESIONS
Cancer Research Institute Siberian Branch of the RAMS
5-th Kooperatiny street, Tomsk, 634050, Russia
e-mail: neg@oncology.tomsk.ru

High prevalence of human papillomavirus (HPV) infection, its causative role in the development of cervical cancer and heterogeneity of HPV types indicate that HPV infection is not only medicobiological problem but also has a social significance. Currently, an HPV DNA test is recommended by international organizations (ASCCP, EUROGIN, IARC WHO) for applying in population screening. For virus-positive women with cervical lesions who are at risk for cervical cancer, follow-up is of great importance, allowing the treatment efficiency to be assessed and the disease outcome to be predicted.

A total of 293 women treated at Tomsk Cancer Research Institute were examined. A median age of the patients was 35.9±0.6 years (range 16 to 80). All patients were divided into 3 groups: 88 patients with benign cervical lesions (30.0%), 101 patients with cervical intraepithelial neoplasia (CIN) (34.3%) and 104 patients with cervical cancer (35.6%). Cervical lesions were histologicaly proven.

HPV screening, differentiation of 12 high-risk HPV types and determination of their concentration were performed using real time multiplex PCR assay. HPV testing was carried out on "Rotor-Gene 6000" amplificatory-machine ("Corbett Research", Australia).

Out of the 293 examined women the first test detected HPV infection in 61.3% of patients with benign cervical lesions, in 72.2% of patients with cervical intraepithelial neoplasia and in 67.3% of cervical cancer patients. HPV type 16 had the highest incidence rate (45.0%) followed by HPV 31–17,0%, HPV 56/33–15,0%, HPV 51/18/52–13,0%, HPV 58/35/39/45–7,0%, HPV 59–5,0%. When studying the virus concentration among HPV-infected women, it was shown that high viral load was observed in 70.9% of patients with benign cervical lesions, in 83.5% of patients with CIN and in 85.7% of patients with cervical cancer.

The median follow-up for 35 patients with cervical lesions was 6.4 months (range 3 to 13). First HPV testing showed that 80.0% of cases were HPV-positive. Patients with benign cervical lesions and CIN were treated in accordance with specific and antiviral program. The second test revealed elimination of HPV infection in 55.6% of primarily HPV-positive patients.

Persistent infection was detected in 35.7% of primarily HPV-positive cases (10 out of 28 patients), mainly in cervical cancer patients. Total number of primarily HPV-positive and HPV-negative patients with cervical cancer was 95.0% and 5.0%, respectively. The corresponding values after the complex treatment were 35.0% and 65.0%, respectively, pointing to the treatment efficiency.

Patients with persistent infection had a high number (83.2%) of cocktail infection cases (combination of several types of papillomavirus) where HPV16 (100.0%) or combination of HPV16 and HPV18 (33.3%) were the most common types. Disease progression was found in 8 out of 35 patients although the most of those cases were HP-free according the second testing (62.5%) that likely to be related to ablation of the HPV-infected tissues.